\documentclass[structabstract]{aa}  
    
\usepackage{natbib,twoopt}
 \usepackage[breaklinks=true]{hyperref} 
 \bibpunct{(}{)}{;}{a}{}{,}    
 \newcommandtwoopt{\citeads}[3][][]{\href{http://adsabs.harvard.edu/abs/#3}%
                                        {\citealp[#1][#2]{#3}}}
 \newcommandtwoopt{\citepads}[3][][]{\href{http://adsabs.harvard.edu/abs/#3}%
                                        {\citep[#1][#2]{#3}}}
 \newcommandtwoopt{\citetads}[3][][]{\href{http://adsabs.harvard.edu/abs/#3}%
                                        {\citet[#1][#2]{#3}}}
 \newcommandtwoopt{\citeyearads}[3][][]%
   {\href{http://adsabs.harvard.edu/abs/#3}{\citeyear[#1][#2]{#3}}}
\usepackage{graphicx}
\usepackage{txfonts}
\begin{document}
   \title{Resolving HD~100546 disc in the mid-infrared: \\
   Small inner disc and asymmetry near the gap.\thanks{Based on observations collected at the European Organisation for Astronomical Research in the Southern Hemisphere, Chile; Guaranteed Time VLTI/MIDI observations: 060.A-9224(A), 074.C-0552(A), 076.C-0252(C) and 076.C-0252(F).}}




   \author{Pani\'c, O. \inst{1,2}
          \and
          Ratzka, Th. \inst{3}
          \and 
          Mulders, G.~D. \inst{4}
          \and
          Dominik, C. \inst{5,6}
          \and
         van Boekel, R. \inst{7}
          \and
          Henning, Th. \inst{7}
          \and
          Jaffe, W. \inst{8}
	\and
          Min, M. \inst{5}          
          }

   \institute{\inst{1}Institute of Astronomy, Madingley Road, Cambridge, CB3 0HA, United Kingdom \\\email{opanic@ast.cam.ac.uk} \\
   \inst{2}European Southern Observatory, Karl Schwarzschild Strasse 2, D-85748 Garching, Germany \\
   \inst{3}Universitaets-Sternwarte Muenchen, Ludwig-Maximilians-Universitaet, Scheinerstr. 1, 81679 Muenchen, Germany \\
   \inst{4}Lunar and Planetary Laboratory, The University of Arizona, 1629 E. University Blvd., Tucson, AZ 85721, USA \\
   \inst{5}Astronomical Institute Anton Pannekoek, University of Amsterdam, Science Park 904,1098 XH Amsterdam, The Netherlands \\
   \inst{6}Department of Astrophysics / IMAPP, Radboud University Nijmegen, P.O. Box  9010, 6500 GL Nijmegen, the Netherlands \\
   \inst{7}Max-Planck Institute for Astronomy, Koenigstuhl 17, 69117 Heidelberg, Germany \\
   \inst{8}Leiden Observatory, Leiden University, Niels Bohrweg 2, 2333 CA Leiden, The Netherlands \\
         }

   \date{Received 14 March 2012; Accepted 29 November 2013}

 
  \abstract
   {A region of roughly half of the solar system scale around the star HD~100546 is known to be largely cleared of gas and dust, in contrast to the outer disc that extends to about 400~AU. However, some material is observed in the immediate vicinity of the star, called the inner disc. Studying the structure of the inner and the outer disc is a first step to establishing the origin of the gap between them and possibly link it to the presence of planets.}
   {We answer the question of how the dust is distributed within and outside the gap, and constrain the disc geometry.}
   {To discern the inner from the outer disc, we used the VLTI interferometer instrument MIDI\thanks{The Mid-infrared Interferometric Instrument of the European Southern Observatory's Very Large Telescope Interferometer.} to observe the disc in the mid-infrared wavelength regime where disc emission dominates in the total flux. Our observations exploited the full potential of MIDI, with an effective combination of baselines of the VLTI 1.8~m and of 8.2~m telescopes. With baseline lengths of 40~m, our long baseline observations are sensitive to the inner few AU from the star, and we combined them with observations at shorter, 15~m baselines, to probe emission beyond the gap at up to 20~AU from the star. We modelled the mid-infrared emission using radial temperature profiles, informed by prior works on this well-studied disc. The model is composed of infinitesimal concentric annuli emitting as black bodies, and it has distinct inner and outer disc components.
}
   {Using this model to simulate our MIDI observations, we derived an upper limit of 0.7~AU for the radial size of the inner disc, from our longest baseline data. This small dusty disc is separated from the edge of the outer disc by a large, $\approx$10~AU wide gap. Our short baseline data place a bright ring of emission at 11$\pm$1~AU. This is consistent with prior observations of the transition region between the gap and the outer disc, known as the disc wall. The inclination and position angle are constrained by our data to $i=$53$\pm$8$\degr$ and $PA=$145$\pm$5$\degr$. These values are close to known estimates of the rim and disc geometry and suggest co-planarity. Signatures of brightness asymmetry are seen in both short and long baseline data, unequivocally discernible from any atmospheric or instrumental effects.}
   {Mid-infrared brightness is seen to be distributed asymmetrically in the vicinity of the gap, as detected in both short and long baseline data. The origin of the asymmetry is consistent with the bright disc wall, which we find to be 1-2~AU wide. The gap is cleared of micron-sized dust, but we cannot rule out the presence of larger particles and/or perturbing bodies.}

\titlerunning{Asymmetries and variability around the gap in HD~100546 disc}
\authorrunning{O. Pani\'c et al.}
   \maketitle
%

\section{Introduction}

Gaps and holes are the most prominent and observable features that a planet may cause in circumstellar discs \citep[e.g.,][]{paardekooper2004,crida2006}. 
These features are observed in a number of young discs, either with direct imaging in the (sub)millimetre  and infrared \citep{pietu2006,brown2009,thalmann2010,andrews2011,quanz2011} or indirectly from the observed deficit in the mid-infrared range of the broad-band spectral energy distribution \citep{calvet2002,bouwman2003,brown2007}. Photo-evaporation or substellar companions are possible physical processes responsible for gap formation. 
Although the dynamical influence of embedded bodies has a limited spatial range, gaps and holes in the inner disc can have a major impact on the fraction of stellar light that reaches the outer portions of the disc. Stellar illumination governs the large scale temperature and vertical structure of a disc, and these are, in turn, important parameters in dust evolution and radial migration. Gap structure is therefore crucial both as a direct consequence of orbiting planets and an influential factor for the planet formation process. 

A number of known discs around nearby (100-200~pc) young stars possess both substantial inner clearings ($\approx$~10~AU) and disc material near the star \citep[e.g.,][]{espaillat2010}. The aim of our observations is to spatially resolve dust emission on solar-system scales from a bright showcase example of these discs, and provide a geometrical characterisation of its gap. 
The target of our study is the nearby star HD~100546 \citep[97$\pm$6~pc;][]{vanleeuwen2007}.  
This is one of few prominent objects that have driven the research of protoplanetary discs and the link with planet formation over the past decade. The disc around this star has a gap in gas and dust distribution inwards from approximately 10 - 13~AU. This is confirmed by ample observational evidence including both SED modelling and direct observations \citep{bouwman2003,benisty2010,tatulli2011,mulders2011,grady2005a,acke2006,vanderplas2009}. 

The large outer disc has been imaged using scattered light observations \citep{pantin2000,augereau2001,grady2005,ardila2007,quillen2006,boccaletti2013}, and it exhibits a multiple spiral arm structure. The disc has a significant gas reservoir and emits strong low- and high-excitation CO emission at radii well beyond the inner few tens of AU \citep{panic2010,goto2012}. Because of this, and especially because of its strong excess emission, the disc around HD~100546 cannot be considered as a debris disc but rather as a hydrostatically supported viscous gaseous disc similar to discs commonly seen around much younger low- and intermediate-mass stars. 

With an age of more than 5~Myr \citep{vandenancker1997,guimaraes2006}, well beyond the average dispersal timescale in discs \citep{haisch2001,hillenbrand2008}, and with the planet-forming regions severely depleted of material, this disc is more likely to have already formed planets rather than being an active site of ongoing planet formation. \citet{quanz2013} detect an emission source at a projected distance of 68~AU from the star, and favour an interpretation with a planet of 15-20~M$_{Jupiter}$.
Putative planets in the disc of HD~100546 would have already dynamically `sculpted' their physical environment and thus modified or erased the initial conditions in which they had been formed.  We therefore regard this disc primarily as a laboratory for studying disc-planet interactions.

Study of the inner disc at near-infrared wavelengths allows us to locate and characterise the inner rim of the disc, the region closest to the star. In \citet{benisty2010,tatulli2011} detailed studies of the inner rim were carried out, placing its location at 0.24~AU. These observations and other prior SED modelling do not, however, provide information on the distribution of colder dust present directly behind the hot rim or inside the gap. To investigate the spatial distribution of the dust in the inner few tens of AU in the disc, we conduct observations of the thermal mid-infrared emission. Mid-infrared emission from HD~100546 is bright and extended, as shown by \citet{verhoeffPhD}, \citet{liu2003}, \citet{leinert2004} and \citet{vanboekel2004}. 
We used the Very Large Telescope Interferometer \citep[VLTI,][]{glindemann2003}, and its Mid-Infrared Interferometric Instrument \citep[MIDI,][]{leinert2003a} to observe HD~100546. MIDI provides interferometric data in the spectral range 8-13~$\mu$m, and is used to combine the signal from pairs of 8.2~m unit telescopes (UTs) and 1.8~m auxiliary telescopes (ATs). With former baselines of 47-130~m, the spatial scales of about 10~mas are efficiently probed. The latter baselines offer more flexibility in the range from 10~m to 200~m.
Observations of discs with MIDI on the UT baselines in the past have characterised the inner discs around several young stars \citep[e.g.,][]{leinert2004,ratzka2007, difolco2009}. The possibility of combining these existing data with those taken on shorter baselines, as in the present work, offers the potential to further improve our understanding of the inner disc regions and obtain complementary information about the structure of discs on larger scales. In \citet{mulders2013} we interpret these MIDI data in the context of a rounded disc wall dynamically sculpted by a planet and suggest that there is a possible planet with 20-80~M$_{Jupiter}$ within the gap.

In Sect.~\ref{datared} we present our interferometric MIDI data and the data reduction methods employed, with special attention to chromatic phase measurements. In Sect.~\ref{results} we assess the data quality and present results regarding the spatial scales and physical regimes probed by these observations. 
In Sect.~\ref{discussion} we describe our modelling approach, the code, and model parameters. In Sect.~\ref{modellingresults} we present our model of the mid-infrared emitting regions of HD~100546, consisting of the inner and the outer discs. We discuss the constraints on the inner disc size and investigate the origin of the correlated flux variations seen on the 41~m baselines. The symmetry of the inner and outer disc are assessed using the chromatic phase measurements. Section~\ref{summary} summarises our conclusions.

\begin{table*}
\caption{Journal of MIDI observations of the science target HD~100546 and the associated calibrators. Telescopes U2 and U3 are 8.2 m UTs, and E0 and G0 are 1.8 m ATs. Dispersers used are indicated by (g) for grism and (p) for prism. Fluxes are included for the calibrators.}
\label{tab1}
\centering
\begin{tabular}{llllll}
\hline\hline
Telescopes & Date & Universal time & Target & \multicolumn{2}{c}{Projected baseline} \\
\hline
\noalign{\smallskip}
U2-U3 (p) & 2004 June 03 & 01:24 - 01:32 & HD~100546 & \ 34.8~m & \ 74.2$\degr$ \\
U2-U3 (p) & 2004 June 03 & 01:49 - 01:57 & HD~120404 (9.6~Jy) & \ 36.4~m & \ 66.3$\degr$ \\
U2-U3 (p) & 2004 June 03 & 03:46 - 03:58 & HD~169916 (21.6~Jy) & \ 44.6~m & \ 18.3$\degr$ \\
\hline
U2-U3 (p) & 2004 December 28 & 07:45-07:56 & HD~98292 & \ 42.1~m & \ 28.8$\degr$ \\
U2-U3 (p) & 2004 December 28 & 08:09-08:19 & HD~100546 & \ 41.3~m & \ 30.8$\degr$ \\
U2-U3 (p) & 2004 December 28 & 08:38-08:49 & HD~102461 & \ 44.4~m & \ 31.1$\degr$ \\
 \hline
U2-U3 (g) & 2005 December 27 & 07:31-07:48 & HD~112985 & \ 42.0~m & \ 5.0$\degr$ \\
U2-U3 (g) & 2005 December 27 & 08:09-08:24 & HD~100546 & \ 41.4~m & \ 29.5$\degr$ \\
U2-U3 (g) & 2005 December 27 & 08:43-09:00 & HD~98292 & \ 41.1~m & \ 39.0$\degr$ \\
 \hline
E0-G0 (p) & 2006 February 13 & 03:36-03:48 & HD~89388 & \ 16.0~m & \ 50.7$\degr$ \\
E0-G0 (p) & 2006 February 13 & 04:01-04:11 & HD~100546 & \ 16.0~m & \ 39.2$\degr$ \\
E0-G0 (p) & 2006 February 13 & 04:52-05:02 & HD~82668 & \ 15.7~m & \ 76.1$\degr$ \\
 \hline
E0-G0 (p) & 2006 February 16 & 05:21 - 05:34 & HD~89388  & \ 15.7~m & \ 76.0$\degr$ \\
E0-G0 (p) & 2006 February 16 & 05:48 - 06:00 & HD~100546 & \ 15.8~m & \ 66.6$\degr$ \\
\hline
E0-G0 (p) & 2006 February 16 & 07:40 - 07:53 & HD~89388 & \ 14.1~m & \ 106.9$\degr$ \\
E0-G0 (p) & 2006 February 16 & 08:08 - 08:20 & HD~100546 & \ 14.9~m & \ 99.3$\degr$ \\
\noalign{\smallskip}
\hline
\end{tabular}
\end{table*}

\section{Observations and data reduction}\label{datared}

\begin{figure}
   \centering
   \includegraphics[width=8.cm,angle=0]{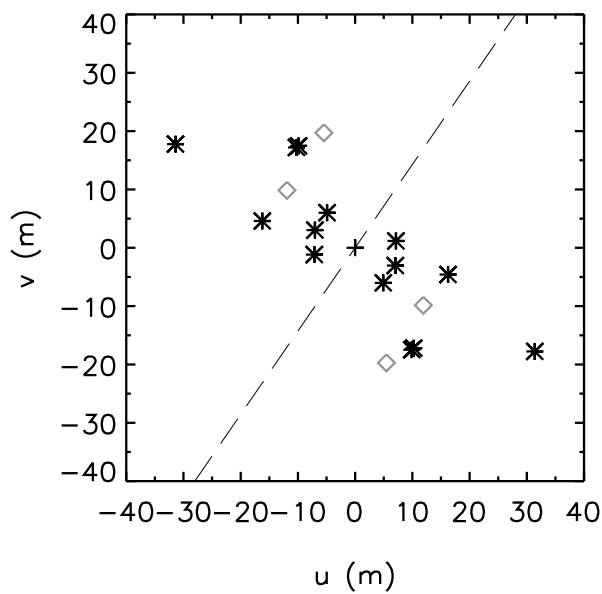}
   \caption{uv-plane coverage of all MIDI observations HD~100546 taken before 2012. North is up and East is left. The baselines presented and analysed in this paper are marked by black asterisks. The longest baseline is presented in \citep{leinert2004}. In addition to their analysis we here present the chromatic phases from this measurement (see Sect.~\ref{asym}). The data from the remaining baselines, shown with grey diamonds, are excluded because of poor quality. The orientation of disc major axis used in this work is shown with the dashed line.}
      \label{uvplot}
    \end{figure}

HD 100546 was observed in several campaigns with MIDI at the Very Large Telescope Interferometer (VLTI). For a detailed description of the instrument and its operation, see \citet{leinert2003a}, \citet{leinert2003b}, \citet{morel2004}, and \citet{ratzka2005}. Both the 8-metre UTs and the 1.8-metre ATs have been  used. A detailed journal of observations is given in Table~\ref{tab1}. Figure~\ref{uvplot} depicts the lengths and orientations of all baselines on which HD~100546 has been observed with the VLTI so far, compared to the orientation of the major axis of the disc, as derived in Sect.~\ref{outerdisc}.

The MIDI spectrometer can be configured with either a prism with $\lambda/\Delta\lambda\simeq 30$ or
a grism with $\lambda/\Delta\lambda\simeq 230$.  The prism is somewhat more sensitive and was used for
all observations except those in 2005.  The grism observations in 2005 do not show
any narrow spectral features that would be unresolved by the prism.

\subsection{Observing procedures}

The interferometric measurements consisted of sequences of 8000 to 12000 frames. The exposure time per frame was 12-22 milliseconds for prism measurements, and 36 milliseconds for grism measurements.  This time is chosen to prevent detector saturation by the bright mid-infrared background.  The optical path difference (OPD) between telescopes was varied in a cyclical manner by $\sim \lambda/4$ per frame with a piezoelectric mirror in order to modulate the interferometric signal. This allows much of the slowly varying sky background signal to be removed by high-pass filtering.

In our measurements, we have not employed a hardware spatial filter and did not
measure simultaneous non-interferometric (`photometric') fluxes. Instead, the photometric measurements are taken after the interferometric observations.
Thus our results are expressed as the {\it correlated flux} between 
the telescope pairs, rather than as {\it visibilities}, i.e., correlated 
fluxes/photometric fluxes. For more details on the correlated flux mode, see the MIDI User Manual. This procedure was followed because
the photometric fluxes are inaccurate owing to the high background 
fluctuations at mid-infrared wavelengths. Moreover,
the influence of seeing fluctuations on the correlated fluxes is 
relatively weak because the adaptive optics hardware on 8-m telescopes 
makes them essentially diffraction-limited at these wavelengths.  
Practice with MIDI has shown that, with the exception of the very brightest sources, the correlated fluxes are more stable and accurate than visibilities.  
In our experiments, additional non-simultaneous photometric fluxes were in fact measured to monitor overall flux changes.

\subsection{Calibration}

Calibration measurements to remove the effects of atmospheric variations were carried out typically
30 minutes before or after the target exposures on calibrators near on the sky.
The calibrators are listed in Table \ref{tab1} and have a flux of 5~Jy or more at 10$\,\mu$m. They have no associated circumstellar emission at mid-infrared wavelengths, no known close companions, and no significant variability, and they are not significantly resolved at 100-metre baselines at 10~$\mu$m.
We selected several calibrators from the MIDI Calibrator Catalogue\footnote{http://www.eso.org/$\sim$arichich/download/vlticalibs-ws/} and some from CalVin\footnote{http://www.eso.org/observing/etc/}.
A subsample of these calibrators, as indicated in Table \ref{tab1}, was used for absolute flux calibration\footnote{http://www.eso.org/sci/facilities/paranal/instruments/midi/tools/}. Detailed photometric spectra of these calibrators and estimates of their diameters are taken from a database of R. van Boekel based ultimately on infrared templates created by \citet{cohen1999}.  The details of the construction of the photometric spectra are described in \citet{vanboekelPhD} and  \citet{verhoelstPhD}.

Calibration uncertainties introduce uncertainties to the derived fluxes in addition to those arising from noise.  In our plots of correlated flux (Figures~\ref{corrflux} and \ref{vari2}), the noise uncertainties are indicated by the error bars. The magnitude of the calibration uncertainties,which move
the whole spectra up or down, is indicated by the curves that represent the results derived by two different calibrators.  Generally, the overall calibration uncertainty of UT observations is $\sim 10$\% and of the AT observations $\sim 20$\%.

The observing sequence described above provides data affected by atmospheric and instrumental correlation losses. To correct for this, the sequence is repeated for a source with known diameter. Ideally, the calibrator observation is close to the target observation in time and in position on the sky, so that the instrumental and atmospheric effects are similar. The instrumental setup, e.g. spectral resolutions, must be identical.

\subsection{Data reduction}

For the data reduction, we used Version 2.0 of {\tt MIA+EWS}\footnote{http://www.strw.leidenuniv.nl/$\sim$nevec/MIDI/}$^,$\footnote{http://www.mpia-hd.mpg.de/MIDISOFT/}. This version is distinguished from the previous ones by an increase in sensitivity due to use of a more sophisticated model of atmospheric time variations in the OPD. Also, there is a reduction of noise bias in weak correlated fluxes. 

We chose the {\tt EWS} branch, which is based on a shift-and-add algorithm in the complex plane. In this algorithm, the atmospheric group OPD is estimated for each plane using all frames within one atmospheric correlation time. The phases of the complex visibility of that frame A($\lambda$)$\cdot$e$^{i\phi(\lambda)}$ are shifted to correct for this OPD.  Then all such shifted frames are averaged.
Thus, this algorithm is `coherent' \citep{jaffe2004, ratzka2009}.
The other branch of the data reduction software, called {\tt MIA}, analyses the power spectrum. The description of the data reduction steps of {\tt MIA} can be found in \cite{leinert2004} and \cite{ratzka2005}. Generally {\tt EWS} is more accurate for weak correlated fluxes. Here we use {\tt MIA} only for an additional consistency check of the {\tt EWS} results.

\subsection{Chromatic phases}

The coherent reduction method used here allows the determination of phase changes with wavelength, otherwise known as differential or chromatic phases.  The signal measured by MIDI at wave number $k=2\pi/\lambda$ at a specific time can be written as $\hat A=A(k)\exp(\theta(k)+kD_a(t)+\phi_a(t))$, where $A(k)$ is the source correlated flux amplitude, $\theta(k)$ the source phase, and $\phi_a(t)$ and $D_a(t)$ the unknown atmospheric phase and delay.  We assume for now that $\phi_a$ and $D_a$ are not wavelength dependent (see below). At each time, we can fit a linear function of $k$ to the phase of $\hat A$ to determine and remove $D_a$ and $\phi_a$ from the measurements.  Unfortunately, this destroys some information about the source phase $\theta$.  If we write $\theta(k)=\theta_0+\theta_1 k+\theta_c(k)$, we see that the delay fitting procedure will also remove $\theta_0$ and $\theta_1$ from the data, leaving only the {\it chromatic phase} $\theta_c$ containing second and higher order variations with $k$.  This removal of the constant and first-order phases means that the measured complex correlated fluxes cannot be used directly in a Fourier transform reconstruction of the source structure, but the chromatic phases can be used to constrain source modes, analogous to the use of closure phases in optical and radio interferometry.

\subsection{Atmospheric dispersion}

In the above discussion we assumed that $D_a$ and $\phi_a$ do not vary with wave number $k$.  This is of course not strictly true, because the dispersion of the
atmospheric refraction is determined by the chemical constitution of the air.
Detailed models for the dispersion have been calculated by \cite{mathar2007} and
other authors. For air conditions similar to that at the VLT, the average refractivity ($n-1$) in the N-band is approximately $10^{-4}$. The dispersion is chiefly due to water vapour and is approximately proportional to wavelength.  If there is a difference in atmospheric path between the two telescopes, the average refractivity contributes the linear term $kD_a$ in the measured phase and the water vapour dispersion to the constant term $\phi_a$. The residual refractivity after removal of these terms is $\sim 10^{-8}$.  Thus the ratio of dispersive to average delay is $\sim 10^{-4}$ and the passage of light in the N-band ($\lambda=10$~$\mu$m) through one metre of air produces dispersive phases of order $10^{-3}$ wavelengths, or 0.3 degrees.

The total atmospheric path difference above the telescopes $D_a$ fluctuates by at most 100~$\mu$m during one source measurement and less than 1~mm between source and calibrator.  The dispersive part of these fluctuations is reduced by a factor of $\sim 10^{-4}$ to $<<0.1\ \mu$m.  Thus the fluctuation contribution to the chromatic phase that is not removed by the calibration process is $<10^{-2}\lambda\simeq 3^{o}$. A potentially larger contribution comes from the VLTI delay lines. These delay lines compensate, in air, for the delay in arrival of the source signal due to the baseline between the two telescopes, which occurs effectively in a vacuum. The refractivity of the air in the delay lines introduces dispersive phase shifts equivalent to passing through a length of air that is of the same order of magnitude as the distance between the telescopes. A delay of $\sim 50$~m would produce chromatic phases of about 15$\degr$, and these are in fact seen in uncalibrated observations.  However, we have chosen our calibrators close on the sky to our target, so that the delay line path differences between target and calibrator are never more than a few metres.  This reduces the residual chromatic phases to a few degrees.

The validity of these calculations is shown by calculating the residual chromatic phases between two calibrator measurements taken one hour apart, i.e., twice the calibrator-target time difference.  The delay line difference between the calibrators is of the same order as that between target and either calibrator.  The residuals, plotted in the bottom panels of Fig.~\ref{phase}, are a few degrees between 8~$\mu$m and 13~$\mu$m wavelengths. These give us confidence that source chromatic phases of more than a few degrees are in fact associated with the source itself and not the atmosphere.

\section{Results}\label{results}

\subsection{The total N-band fluxes}\label{totalflux}   
   
   \begin{figure}
   \centering
   \includegraphics[width=9.cm,angle=0]{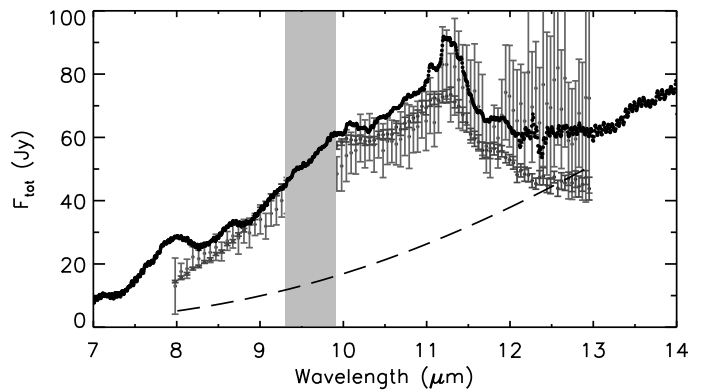}
   \caption{ISO-SWS spectrum from \citet{acke2004} is shown in black, a total calibrated flux measurement with an AT (1.8~m) in light grey and with a UT (8.2~m) in dark grey. The dashed line is the prediction from our model described in Sect.~\ref{outerdisc}. A masked region from 9.3 to 9.9~$\mu$m corresponds to the wavelength range of the atmospheric ozone features affecting our data.}
              \label{flux}
    \end{figure}  

   \begin{figure}
   \centering
   \includegraphics[width=9.cm,angle=0]{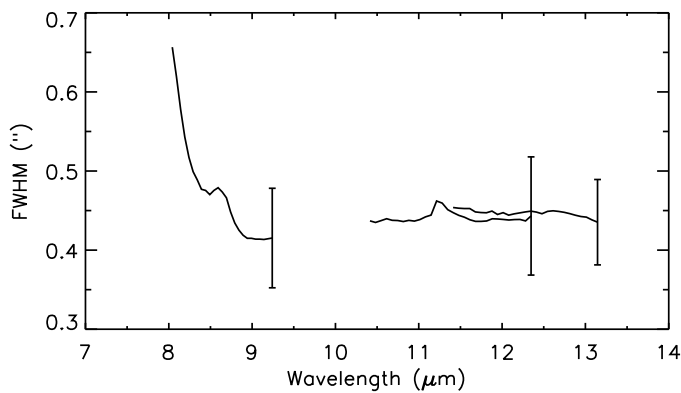}
   \caption{The deconvolved size of HD~100546 as a function of wavelength, derived based on VLT/VISIR observations presented in \citet{verhoeffPhD}. Figure adapted from a sub-plot in Fig.~2.9 of \citet{verhoeffPhD}, Chapter 2. The errors shown at the end of each wavelength setting are medium values.}
              \label{fwhm}
    \end{figure}

Previous mid-infrared observations of HD~100546 have been carried out both with the Infrared Space Observatory (ISO) and with ground-based instruments. In Fig.~\ref{flux} the ISO data \citep{acke2004} can be seen. N-band flux of HD~100546 is clearly dominated by a number of spectral features, including a dominant and broad amorphous silicate feature centred at 9.7~$\mu$m, and somewhat narrower PAH bands atop the silicate feature, at 7.7, 8.6, and 11.2~$\mu$m. The strong peak near 11~$\mu$m is due to a blend of PAH and crystalline silicate features. Thorough analyses of the spectral features towards our target are presented in \citet{bouwman2003},\citet{acke2004}, and \citet{mulders2011}. Figure~2 in \citet{bouwman2003} nicely shows the contributions of different features to the cumulative flux. Our present work focusses on the spatial origin of the emission, and sets aside the complex aspects of dust spectral features, in favour of a more simplistic approach where thermal continuum emission from featureless dust is assumed. In Sect.~\ref{discussion} we return to our analysis approach in detail. 

In addition to our interferometric data, accompanying photometric measurements were taken with the individual 1.8 and 8.2~m telescopes. Calibration uncertainties of these photometric measurements are 20$\%$ and 10$\%$ for the two types of telescopes, respectively. Figure~\ref{flux} shows the total calibrated flux measurements from a 1.8 and a 8.2~m telescope, in comparison with the ISO data. There are some differences, particularly near the PAH 7.7 and 8.6~$\mu$m features, and at wavelengths larger than 10~$\mu$m. Total fluxes shown in Fig.~\ref{flux} are calibrated but not masked, and are therefore only limited by the beam and slit sizes. Because of their larger diameter, the point spread function (PSF) of the UTs is smaller than that of the ATs: 0$\farcs$25 and 2$\farcs$2, respectively. The corresponding slit sizes are 0$\farcs$52 and 2$\farcs$3.  The PSF of the ISO telescope at these wavelengths is $\sim 7$". 
In neither of our telescopes do we completely recover the emission from the aforementioned PAH features.

A reason may be the known large spatial extent of the mid-infrared emission towards our target, as studied in great detail in \citet{verhoeffPhD} using VLT/VISIR observations\footnote{Based on observations collected at the European Organisation for Astronomical Research in the Southern Hemisphere, Chile; Guaranteed Time VLT/VISIR observations 075.C-0540A.1}. Figure \ref{fwhm} shows their derived deconvolved FWHM sizes of HD~100546 in three wavelength settings across the N-band. All PAH features are extended, and in particular, the FWHM of the 7.7~$\mu$m feature exceeds the slit size of a UT significantly. 
Similarly, in their comparison of the total flux measured with a 8.2~m UT to measurement with TIMMI2, \citet[][their Fig.~2]{leinert2004} show that a large portion of the 8-9~$\mu$m flux is not recovered with a UT. The large FWHM of roughly 150~AU was found by \citet{vanboekel2004} for the PAH emission from HD~100546.
Another difference between the ISO and our ground-based data is clearly seen at longer wavelengths. In comparison to the data from a 8.2~m UT telescope, a fraction of the long wavelength flux, originating beyond about 25~AU from the star, is blocked by the slit of 0$\farcs$52.

\subsection{Interferometric data}\label{intf}

Another effect on the angular scales accessible with our observations is introduced by our choice of spatial masks used in the MIDI data reduction. We used the standard masks of the EWS software. This was done to suppress unwanted background emission, and is essential in observations of weak sources. This limits the effective field of view to $\sim 0.4\farcs$ on the UTs and $\sim 1.6\farcs$ on the ATs. 
The comparison between the masked and unmasked N band fluxes measured using UTs shows that masking blocked out less than 10\% of the flux. For the ATs, the effect is even larger, 25$\%$ to 30$\%$. Our final interferometric data - the correlated fluxes - are therefore derived from a brightness distribution different than the total flux seen with the individual telescopes. We see in Sect.~\ref{totalflux} that even the total fluxes are affected by the slit size for our bright and extended source. Because of these difficulties, we chose to carry out a comparative analysis of the total and correlated fluxes, instead of the commonly used approach of modelling the visibilities ($F_\mathrm{corr}/F_\mathrm{tot}$) directly. 

The interferometric field of view\footnote{The interferometric field of view equals $R*\lambda/B$, with $R$ the spectral resolution and $B$ the baseline length.} defines the largest angle 
on the sky from which coherent light can be combined. For the prism ($R=$30) observations on 41~m baselines, it ranges from 1$\farcs$2 to 1$\farcs$8 over the N-band wavelengths, while for grism ($R=$230) observations it ranges from  9$\farcs$1 to 13$\farcs$7. 
On the shorter, 15~m AT baselines (prism), we obtain interferometric fields of view of 3 - 6$\arcsec$. The angular limitations imposed by the interferometric field of view are therefore not relevant for our data, because these angular scales are much larger than those delimited by the photometric field of view, slit size, and masking. 

\begin{figure*}
   \centering
   \includegraphics[width=18cm]{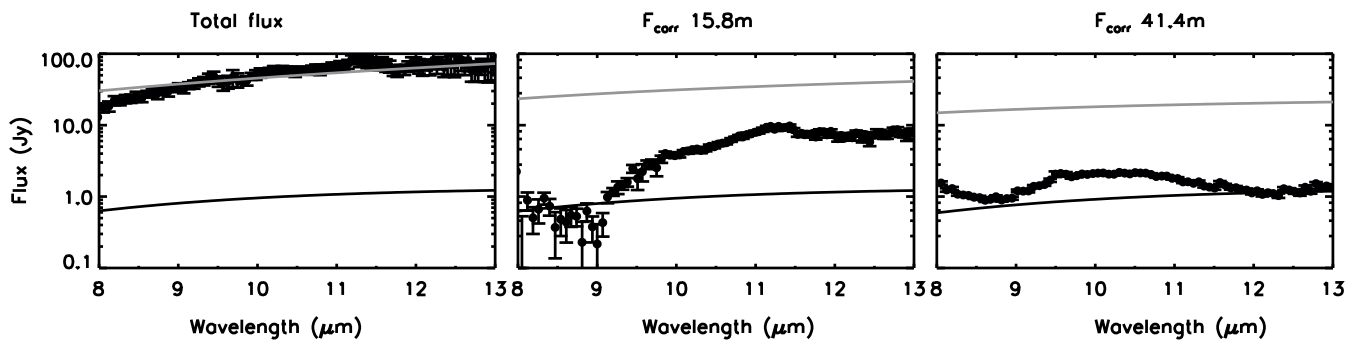}
    \caption{Left to right: The data shown correspond to total fluxes, correlated fluxes on a 15~m baseline, correlated fluxes on a 40~m baseline. Neither of the radially continuous disc models with radii 0.7 and 25~AU (black and grey lines, respectively) match all three types of data.}\label{gap} 
    \end{figure*}

Figure~\ref{gap} allows an order-of-magnitude comparison between the total fluxes and the fluxes correlated on our long and short baselines. The correlated fluxes are just a minor fraction of the total flux, and in the case of the long baselines, the 1-2~Jy levels seen on the long baselines are comparable to or smaller than the 10$\%$ errors on the total fluxes, which are largely dominated by calibration uncertainties. That the total and correlated fluxes differ by this much cannot be explained by the way the interferometric observations filter out emission on larger scales than the preferential angular scale prescribed by the baseline length. In Sect.~\ref{modellingresults} we show that, in concordance with the wealth of prior observational evidence, the small fraction of flux correlated on the long baselines is due to the presence of a large gap.

\begin{figure}
   \centering
   \includegraphics[width=8cm]{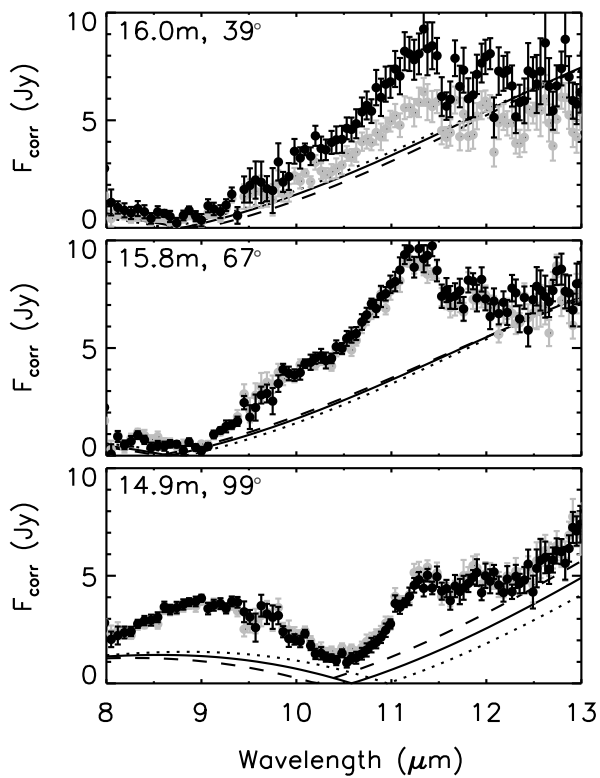}
    \caption{The correlated fluxes obtained with MIDI using the ATs in February 2006. Baseline lengths and orientations are indicated. Black and grey symbols correspond to data reduction using two different calibrators, observed immediately prior to or following the science target. The calibration errors are within 20$\%$. The plotted data errors are instrumental errors. Full lines show the model prediction for the correlated fluxes on these baselines (see Sect.~\ref{outerdisc}). The dotted and the dashed lines present the model result for position angles 140$\degr$ and 150$\degr$, respectively, compared to our model with 145$\degr$ shown with full lines.}
              \label{corrflux}
    \end{figure}

Figures~\ref{corrflux} and \ref{vari2} show the correlated fluxes $F_\mathrm{corr}$ on all our baselines.  We focussed on those portions of the mid-infrared spectrum where both the total and correlated fluxes are most likely dominated by the continuum emission, and not spectral features. These regions are 8-9~$\mu$m and 12-13~$\mu$m. 
The correlated fluxes measured on the UT baselines (Fig.~\ref{vari2}) are not only quantitatively, but also qualitatively different from those seen on the AT baselines (Fig.~\ref{corrflux}). The UT correlated fluxes are comparable across the N-band spectral range, suggesting that the bulk of this emission arises from relatively hot regions, above a few hundred K. Such temperatures are found in the inner few AU from the star. The AT correlated fluxes, on the other hand, increase towards longer wavelengths, indicating that the dominant emission on these baselines comes from relatively colder material of 200-300~K. These temperatures are found in the outer disc, well beyond the inner few AU.

The spectral shapes of the correlated fluxes measured on the UT baselines are significantly different from one another (Fig.~\ref{vari2}, upper panels). In particular, there are two baselines with the same length and orientation with significantly different shapes. The only instrumental difference between these two measurements is the spectral resolution (prism vs. grism). The observed variability cannot be related to this in any way, because the variation is present over a spectral range that is much larger than the channel spacing. The variability is significantly larger than the atmospheric effects, statistical, instrumental, and calibration errors and cannot arise from any instrumental effects. Therefore it is without any doubt caused by a change in the spatial distribution of N-band brightness toward our science target on the different dates of observation. In comparison to these two baselines, the remaining UT baseline of different geometry has much lower correlated fluxes overall.

\subsection{Dust spectral features}\label{dustspec}

\begin{figure*}
\centering
\includegraphics[width=18cm,angle=0,clip]{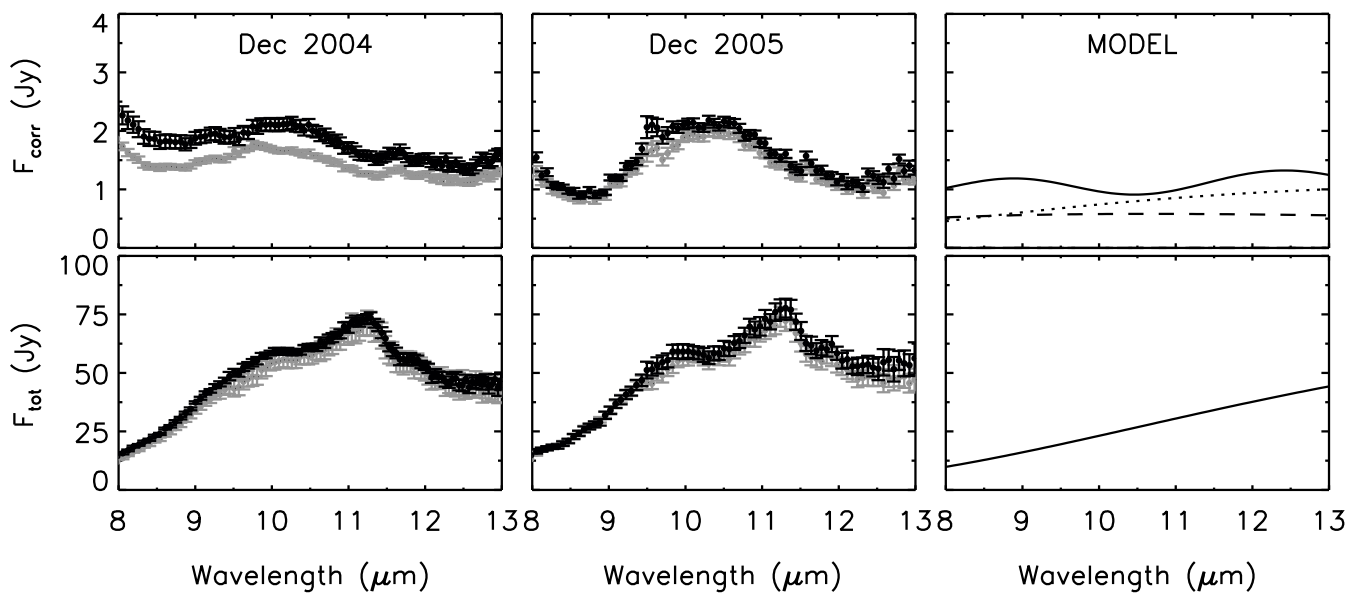}
\caption{{\it Upper panels:} The correlated fluxes from UT-baseline measurements observed in December 2004 and December 2005, and from our model. The baseline length and orientation is 41~m and 30$\degr$ in both observations. Black and grey symbols are used to distinguish results obtained by using two different calibrators. Error bars correspond to instrumental errors. The dashed and dotted lines show the contribution of the inner rim and the inner disc, respectively (Sect.~\ref{innsize}). The full line shows the final correlated flux including contributions from all disc components (Sect.~\ref{outerdisc}). {\it Lower panels:} Total unmasked fluxes measured on the UTs simultaneously with the above observations from December 2004 and December 2005, and the modelled total flux. As before, black and grey symbols correspond to the use of two different calibrators. Errors shown are due to instrumental errors.} 
\label{vari2}
\end{figure*}

As already mentioned in Sect.~\ref{totalflux}, the total fluxes are dominated by a number of spectral features, some of which are extended and therefore not present in all or some of our interferometric data. The UT data are especially affected by the slit size and lack some of the short-wavelength PAH features. The only suspected feature seen in our UT correlated fluxes, for example, is a broad and weak bump at 9-12~$\mu$m (see Fig.~\ref{vari2}, upper panels), which may or may not be related to the silicate feature seen in the total N-band fluxes. A sharp and narrow ring of emission located beyond a few AU from the star can also cause a frequency-dependent wavy pattern on the 41~m UT baselines. Given the above-mentioned limitations we, did not try to model or analyse the spectral features potentially present in our data.

\section{Modelling approach}\label{discussion}

\subsection{Advantages of a direct analysis of correlated fluxes}

With the set of data that differentially probe the inner and the outer discs, we are able to constrain the size of the inner disc, geometry, and location of the bright wall of the outer disc. With the `disc wall' we hereafter refer to the bright, probably directly illuminated inner edge of the outer disc, located close to 10-13~AU.

Because the purpose of our study is not to analyse the spectral features or their extent, but to model the underlying continuum brightness, we limit our work to the analysis and modelling of the correlated fluxes, while ensuring only a very rough continuum-level fit to the total fluxes that are heavily dominated by spectral features and in part cut out by the telescope slit. It is counter-intuitive and extremely difficult to understand and model the visibilities ($F_\mathrm{corr}/F_\mathrm{tot}$) directly in this particular case, so we do not use this approach. 

We calibrated the correlated fluxes directly (Sect.~\ref{datared}). The correlated fluxes obtained in this way are more precise, since they are not linked to the photometric errors involved in measuring the total flux. A separate fit to the correlated fluxes ensures the most efficient exploitation of the information contained in the data. The analysis is more robust than considering visibilities, and it is eased by the intuitive link between the model parameters, such as temperature, and the correlated flux.

\subsection{Computation of the correlated fluxes}\label{comp}

To analyse our data, we developed an IDL-based code that simulates N-band interferometric observations of thermal continuum emission from disc models described by radial distributions of temperature. The correlated flux produced by a radially thin ring of emission of radius $R$ seen at the distance $d$ and measured with a baseline characterised by the uv-distance $uv$ can be described analytically as 
\begin{equation}
F_{corr}(\lambda)=|f(R)\:J_0(arg)|
\end{equation}
where $J_0$ is the zeroeth order Bessel function of the first kind, and its argument $arg=2\pi\:uv/\lambda\:R/d$, while $f(R)$ is the flux emitted by the ringlet at wavelength $\lambda$. If the source brightness is described by more than one concentric ringlet, which is generally the case, we have
\begin{equation}
F_{corr}(\lambda)=|\sum_{R}(f(R)\times J_0(arg))|
\end{equation}
This relation holds regardless of whether the individual ringlets are contiguous, therefore discs with holes or gaps can be modelled in this way. The total flux is given by $F_{tot}(\lambda)=\sum_{R}f(R)$.

Our model of HD~100546 consists of the inner and the outer disc. These components are subdivided into concentric ringlets. The width of ringlets is tuned to a small enough size to effectively sample the corresponding $J_0$ over the observed wavelength range. To obtain the flux from each ringlet, $f(R)$, black body emission at ringlet temperature $T(R)$ is attenuated with $(1-e^{-\tau})$ and N-band extinction. The dust opacity $\tau$ is either constant, a power law, or allowed to increase radially to simulate the disc wall (explained in later sections). 

The values of $J_0$ oscillate between 1 and -1 as a function of wavelength, and some of the zero values may fall within the observed spectral range, depending on baseline and ringlet geometries. For structures with a strongly continuous radial brightness over the entire range of spatial scales starting from those unresolved by a particular observation to those large enough to be effectively resolved out, the positive and negative contributions in $\sum_{R}(f(R)\times J_0(arg))$ amount to a continuous wavelength dependence of the correlated flux and without zero values. On the other hand, sharp contrasts in the radial brightness may strongly accentuate emission from spatial scales well within the range probed by an observation.
In such a situation, the Bessel functions of ringlets located within a sufficiently narrow radial range from one another will be similar in structure, altogether giving rise to points of discontinuity in the flux, with $F_{corr}(\lambda)=0$. This notion will be taken advantage of in Sect.~\ref{outerdisc}, where our modelling is guided by the fact that a bright, directly illuminated, wall of the outer disc is present in HD~100546 \citep{mulders2011, quanz2011}. Sharp brightness contrasts on scales probed by our AT baselines are thus expected. We therefore chose to model the minima in the correlated fluxes on short baselines by adopting a narrow radial component in the disc structure, consistent with the expectation of a bright disc wall. It is important to note, though, that with poor uv-coverage on single baselines, the presence of minima in correlated fluxes is not sufficient evidence of a bright wall in a disc.

\subsection{Indications of a gap}

In Fig.~\ref{gap} we compare the total and the correlated fluxes. Before adopting a more elaborate brightness distribution, we first verified that different data are indeed dominated by emission from different regions of the disc. We show that the differences between the flux measured photometrically and the fluxes measured interferometrically cannot be explained by simple spatial filtering effects from a simple radially continuous brightness distribution. For this purpose we used a simple power-law temperature profile $T(R/0.24\mathrm{[AU]})^{-0.4}$ with radius $R$ and extending from the inner rim at 0.24~AU \citep[][]{benisty2010,tatulli2011}, to $R_{max}$. For $R_{max}=$25~AU and $T=$400~K (Fig.~\ref{gap}), for example, we roughly reproduce the total flux levels, but at the same time the correlated fluxes are overestimated by almost an order of magnitude. On the other hand, with $R_{max}=$0.7~AU and $T=$235~K (Fig.~\ref{gap}), the correlated fluxes at long baselines are matched, but this corresponds to only a minor fraction of fluxes observed on shorter baselines, and it largely underestimates the total flux, underlining the need for some additional emission on significantly larger scales. That radially continuous models fail to simultaneously match the emission seen on different spatial scales implies that a more complicated geometry is needed. This goes hand in hand with the well established notion that a large gap is present within roughly 10~AU of the star from a wealth of observational data, including imaging. 

\subsection{Our model}

In this section, we proceed by modelling the HD~100546 disc as consisting of an inner and an outer disc. In the present work, we derive the most robust disc parameters from the MIDI data alone and make simplest assumptions possible. Mid-infrared emission is not a reliable density tracer, and we focus on the brightness distribution as an indicator of the size and geometry of our source. Deriving the underlying physical conditions (e.g., opacity, dust size) is beyond the scope of this paper, since to constrain these conditions would require a multiwavelength approach with more than just mid-infrared data. Table~\ref{tab2} summarises our model parameters. Wherever observational constraints exist we fix the relevant parameters or vary them in the range of values consistent with observations. The SED model of the disc is used as a valuable reference for physically consistent disc structure and, in particular, for disc temperature (Sect.~\ref{temperature}). Disc orientation is described by the inclination $i$ and position angle $PA$. The stellar emission in N-band is negligible with respect to the dust emission, and is added as a point source black body of 10000~K temperature and 2~R$_{\odot}$ diameter.

\begin{figure}
   \centering
   \includegraphics[width=9.cm,angle=0]{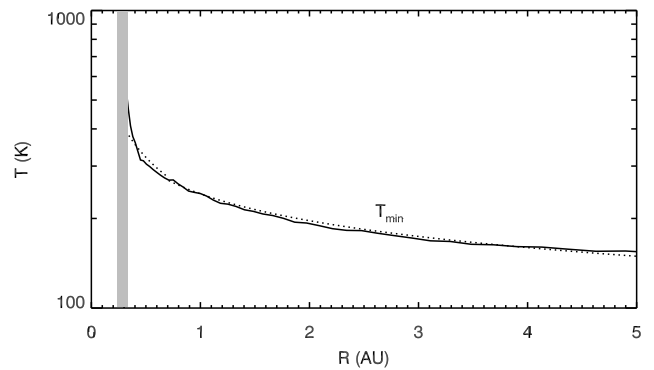}
   \caption{Radial profile of the midplane temperature, $T_{min}$, calculated consistently in case where the $\tau=$1 surface at 10.5~$\mu$m (calculated vertically) coincides with the disc midplane. The dotted line is an analytic fit to $T_{min}$ (see text). The shaded area is the location of the disc rim. }
              \label{temp}%
    \end{figure}

The rim location is set at $R_{in,rim}=$0.24~AU, and the rim is contiguous with the inner disc starting at $R_{in,ID}=$0.34~AU and extending to the free parameter $R_{out,ID}$. The temperature structure of the inner disc $T_{ID}=T_{min}$ is adopted from the SED modelling (see Sect.~\ref{temperature} and Fig.~\ref{temp}). The rim temperature $T_{rim}$ is fit with a power-law slope $q_{rim}$ from the temperature at the inner edge of the inner disc $T_{in,ID}$ inwards. The rim makes a point source contribution on our longest baselines, while it is negligible on our shorter, AT baselines.

Beyond the gap, the outer disc extends from $R_{in,OD}$. We vary this parameter in the neighbourhood of the observationally constrained values 10-13~AU \citep{bouwman2003,quanz2011} to obtain a fit. The size of the disc is fixed to a high enough value  $R_{out,OD}=$25~AU to provide a gradual radial decrease in flux, and does not influence our results. The temperature of the outer disc is a power law $T_{OD}=T_{in,OD}(R/R_{in,OD})^{q_{OD}}$ with the temperature near the gap $T_{in,OD}$ and slope $q_{in,OD}$ as free parameters. We only scan the physically consistent temperature range 200-300~K for $T_{in,OD}$, as expected in the case of an illuminated wall \citep{mulders2011}.

Because we do not intend to derive a detailed physically consistent model of the disc but rather the description of its radial brightness profile, we simplified the calculation of the emission from our models by assuming an optically thick rim and an inner disc with $\tau_{rim}=\tau_{ID}=$1. Such simplification is not possible for the outer disc wall because our observations are particularly sensitive to the contrast between the wall and the gap. To avoid spurious interferometric signal from a high contrast in the flux and to smooth this transition, we introduced $\tau_{OD}$ as a sine function (see Table~\ref{tab2}). The combined effects of the radial increase in optical depth and decrease in temperature gave rise to a relatively smooth peak of the flux behind $R_{in,OD}$. The free parameters $R_{in,OD}$ and $T_{in,OD}$ are degenerate in setting the exact location of this peak, but this location is constrained to 10-13~AU in the literature and we ensure consistency with these constraints.

\begin{table}
\caption{Model parameters. The parameters fitted here are denoted by $\dagger$, the remainder are fixed.}
\label{tab2}
\centering
\begin{tabular}{lll}
\hline\hline
\multicolumn{3}{c}{Geometry}   \\
\hline
$PA$ & 145$\degr\pm$5$\degr$ &  $\dagger$\\
$i$ (0$=$face-on)  &  53$\degr\pm$8$\degr$  & $\dagger$\\
\noalign{\smallskip}
\hline\hline
 & Rim  &  \\
\hline
$R_{in, rim}$   & 0.24~AU  & \\
$T_{rim}=T_{in,ID}(R/R_{in,ID})^{q_{rim}}$ &  &  \\
$T_{in,ID}$ & 369~K  & \\
$q_{rim}$ &  -1.4  &$\dagger$ \\
$\tau_{rim}$  & 1  & \\
\noalign{\smallskip}
\hline\hline
 & Inner disc  &  \\
\hline
$R_{in, ID}$ &  0.34~AU  & \\
$R_{out,ID}$ &  0.7~AU & $\dagger$ \\
$\tau_{ID}$ &  1  & \\
$T_{min}$    &  SED model (Fig.~\ref{temp})  & \\
\noalign{\smallskip}
\hline\hline
 & Outer disc  &  \\
\hline
$R_{in,OD}$ &  9.3~AU & $\dagger$\\
$R_{out,OD}$ & 25~AU & \\
$T_{OD}=T_{in,OD}(R/R_{in,OD})^{q_{OD}}$   &  & \\
$T_{in,OD}$ & 230~K & $\dagger$\\
$q_{OD}$ &  -0.5  &  $\dagger$\\
$\tau_{OD}=\sin{(\pi/2\frac{(r-R_{in,OD})}{(R_{out,OD}-R_{in,OD})})}$ &  &  \\
\noalign{\smallskip}
\hline
\end{tabular}
\end{table}

\subsection{Temperature assumptions}\label{temperature}

Given the strong effect of the disc temperature on the emerging thermal mid-infrared emission, we investigated several physical scenarios and examined the temperature profile of the $\tau=$1 surface at 10.5~$\mu$m (centre of N-band). To make rough estimates of possible temperatures, we used a version of the model of \citet{mulders2011} in which we disregarded the gap, and had a continuous radial distribution of disc material from the dust inner radius at 0.24~AU outwards. In the assumptions of the rim temperature we tried to stay consistent with the findings of \citet{benisty2010} and \citet{tatulli2011}, but acknowledge the strong dependence on the yet unconstrained dust properties. We vary the column density to explore several extreme scenarios, and focus on the innermost several AU. 

We examined a low density case with the 10.5~$\mu$m $\tau=$1 surface as low as the disc midplane ($z=$0), and calculated the temperature structure consistently. In this scenario, the mid-infrared emission arises from the lowest possible temperature $T_{min}$, the disc midplane temperature, shown in Fig.~\ref{temp}.
This is by no means the actual temperature dominating the mid-infrared emission of HD~100546, nor does this physical description correspond to the observed mid-infrared fluxes, but it serves as a valuable indication of how low the temperature in these regions may be. 

\section{Modelling results}\label{modellingresults}

\subsection{Inner disc: The size estimate and variable correlated fluxes}\label{innsize}


The inner disc regions span a temperature range from a few hundred to several hundred Kelvin (see Fig.~\ref{temp}), and emit predominantly in the mid-infrared. The shorter wavelengths are sensitive to even hotter, directly illuminated inner rim of the disc, but such observations \citep[e.g., with AMBER][]{benisty2010,tatulli2011} cannot constrain the distribution of disc material beyond the rim.

Our interferometric data from 40~m UT baselines are particularly sensitive to the spatial scales of less than a few AU. We limit the analysis of those correlated fluxes to placing constraints on the size of the inner disc, and avoid speculating about what may seem like spectral features. The contribution of the thermal emission from a small inner disc to the correlated fluxes is positive and linear. 
If sufficiently bright and narrow, wall of the outer disc produces emission on scales probed by the long baselines. Such features imprint a wavy pattern in the correlated fluxes on the UT baselines, similar to our observations (Fig.~\ref{vari2}), which could easily be mistaken for spectral features or even interpreted in terms of variable dust properties. The correlated flux measurements from Dec 2004 and 2005 are markedly different (Fig.~\ref{vari2}, upper panels). Because of this variability, we focus on fitting the non-variable correlated fluxes at 12-13~$\mu$m, free of known spectral features.

To derive an upper limit on the size of the inner disc, we assume that the mid-infrared flux from this region  is emitted at the lowest temperature found in the disc, $T_{min}$, corresponding to the disc midplane (see Sect.~\ref{temperature})\footnote{This temperature profile can be described analytically as a combination of power-law slopes: $T=$235~K~$(R/{1 \mathrm{AU}})^{-0.42}$  between 0.34 and 0.7~AU and a slightly shallower slope $T=$242~K~$(R/{1 \mathrm{AU}})^{-0.30}$ between 0.7 and 5.0~AU. The temperature from 5~AU to the outer disc wall is roughly 150~K.}. With the remainder of the parameters fixed, we fit the rim temperature with $q_\mathrm{rim}=-$1.4 and obtain an upper limit of $R_\mathrm{out,ID}<$0.7~AU. As explained above, in this instance we ignore the contribution of the outer disc.
The upper right panel of Figure \ref{vari2} shows the resulting contributions from the rim (dashed line) and inner disc (dotted line) to the correlated flux.
Higher temperatures of the rim are possible, but such assumptions need to be compensated for by even smaller inner-disc sizes $R_\mathrm{out,ID}<$0.7~AU. The upper limit of 0.7~AU is robust with respect to the temperature of the emitting layer: at higher temperatures the models yield smaller $R_\mathrm{out,ID}$ to compensate for excess flux. The slope $q_\mathrm{rim}=-$1.4 yields 600~K at $R_\mathrm{in,rim}$. In the extreme case, our UT baseline data can even be fit by a rim alone, with an opportune choice of relatively high temperature, near 1000~K. Even higher temperature of the rim, up to 1750~K is used in \citet{tatulli2011} to fit the H and K-band emission from HD~100546 observed with AMBER and the near-infrared SED, so much smaller inner disc than our upper limit is likely.

We also investigated the case of optically thin N-band emission from the inner disc. For simplicity, we neglected the wavelength dependence of the optical depth. Again, we assumed disc midplane temperatures, but we attenuate the emission by a factor $1-e^{-\tau_0(R_{in,ID}/R)}$ and fit $\tau_0$. We find that in this scenario, the size of the inner disc is not constrained because the flux levels are so low that we no longer can detect any emission beyond 1.5~AU in our observations. A fit to the UT-baseline data is obtained with $\tau_0=$~0.8 in N-band. Lower values of optical depth in N-band are ruled out. It is important to note that the mid-infrared emission is relatively insensitive to large mm and cm-sized particles, and is most sensitive to emission from micron-sized dust. Therefore, any large particles and planetesimals are unconstrained by our data and may be more spatially extended than 0.7~AU.


   \begin{figure*}
   \centering
   \includegraphics[width=18.cm,angle=0]{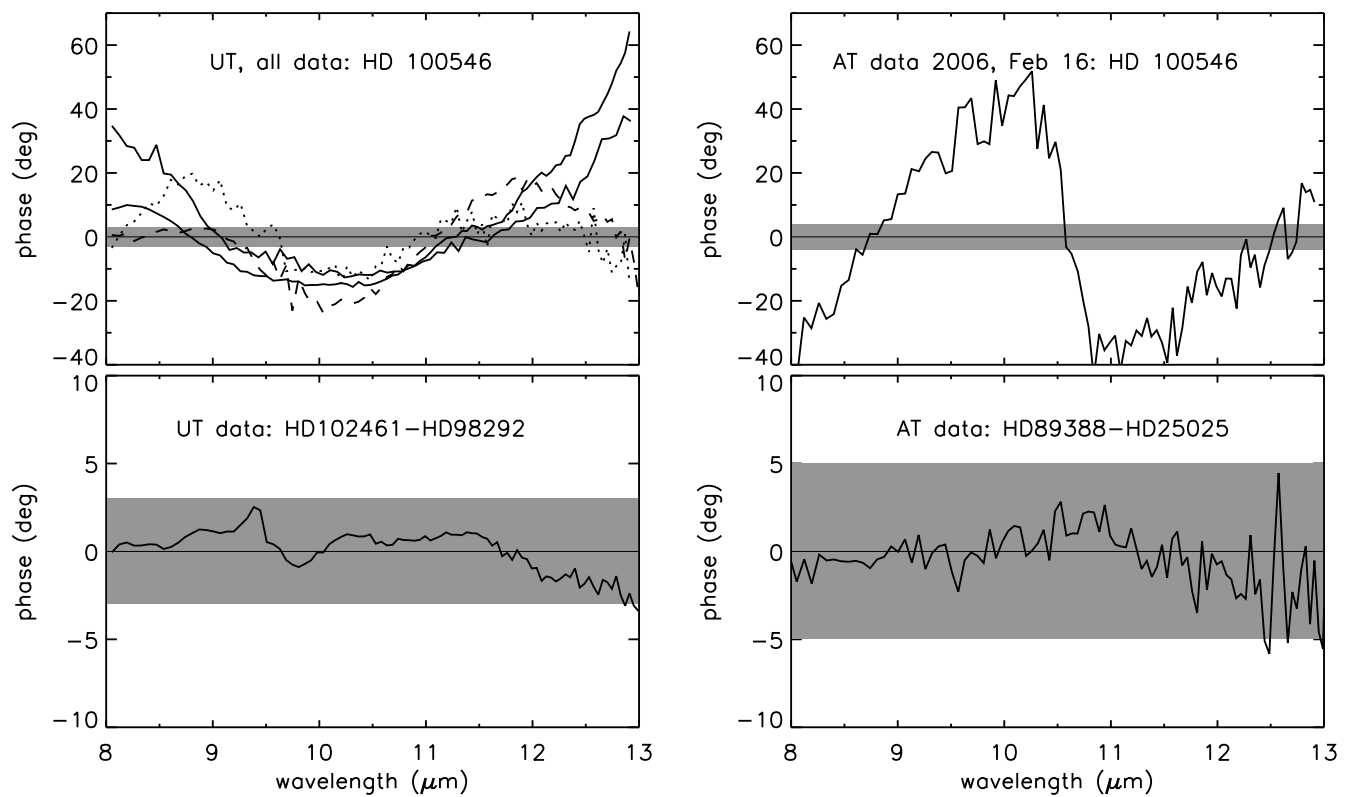}
   \caption{{\it Top left:} Calibrated chromatic phases of our UT baseline measurements of HD~100546. The 34~m baselines is plotted with dotted and 41~m baselines with full lines. The 70~m baseline presented in \citet{leinert2004} is plotted with the dashed line. A linear, atmospheric term has been subtracted from the shown phases. Also shown are the zero phases, as predicted from our point symmetric model. The grey area represents the $\pm$3$\degr$ tolerance for deviation from zero.
{\it Top right:}   Calibrated chromatic phases of our AT baseline measurement of HD~100546 taken on 2006 February 16. The tolerance for deviation from zero (point symmetric model) is $\pm$5$\degr$ and indicated in grey.
{\it Bottom left:} Differences between the measured phases of two point-source calibrators (no asymmetry) at different airmasses, to illustrate the level of atmospheric contributions of our UT measurements.
{\it Bottom right:} As above, for our AT measurements.
}
              \label{phase}
    \end{figure*}
    
As discussed in Sect.~\ref{intf}, there is a significant difference between the two observations obtained on an identical UT baseline (41~m, 30$\degr$). This is shown explicitly in Fig.~\ref{vari2} where correlated fluxes are compared. We establish that no instrumental nor atmospheric effects play a role here, and the difference persists independently of our choice of calibrator. Therefore the variability we observe in the correlated fluxes implies that in December 2004 and 2005, the spatial distribution of the brightness on AU-scales was different. Given the short orbital timescales in the inner disc (less than 1 year at distances less than 1.3~AU from the star), it is plausible that this region undergoes structural changes over a one-year period. Alternatively, the wall of the outer disc, if sharp enough, may contribute to the fluxes probed by our UT baselines; e.g., a change in the brightness ratio between the inner disc and the outer disc wall may have enhanced the wavy pattern of $F_\mathrm{corr}$. Alternatively, rotation of an asymmetry in the inner disc or in the wall of the outer disc can change the brightness distribution, hence $F_\mathrm{corr}$, over time. One-year orbital motion at 11~AU from HD~100546 corresponds to a displacement of approximately 2~AU.

\subsection{Outer disc: Location and width of the disc wall}\label{outerdisc}


As seen in the earlier sections, the amount of flux arising from the inner disc is a minor fraction of the flux measured on the AT baselines, therefore the presence of a small and faint inner disc does not affect the parameters we derive for the outer disc of HD~100546. Instead, our AT data consisting of the strength and spectral shapes of the correlated fluxes across the N-band on three distinct baselines allow us to characterise the outer disc. They probe larger spatial scales where strong emission from extended colder material dominates with about ten times larger fluxes ($\approx$10~Jy vs 1~Jy from the inner disc). In this section, we use the model of the rim and inner disc derived in Sect.~\ref{innsize}, with low, midplane temperature, and $R_\mathrm{out,ID}=$0.7~AU as a starting point and add an outer disc component.

A prominent feature in each correlated flux measurement shown in Fig.~\ref{corrflux} is the presence of a prominent zero-flux minimum. At the end of Sect.~\ref{comp}, the link between the minima and sharp, radially narrow features is explained. Based on solid observational evidence of a bright wall close to 13~AU, possibly a directly illuminated edge of the outer disc, we proceed in our analysis with an outer disc model with strong emission from the wall in the neighbourhood of 13~AU, slightly smoothed to account for optical depth effects (see above), and a gradual decrease in disc brightness with radius, consistent with the deconvolved FWHM of 0$\farcs$45 observed with VLT/VISIR \citep{verhoeffPhD}. 

The presence of a bright wall has effects on the long baseline data by adding a wavy component. Setting a requirement to have comparable correlated fluxes at 8 and at 13~$\mu$m on the 40~m baselines (Fig.~\ref{vari2}), we obtain $R_{in,OD}=$9.3~AU. Starting with this basic radial distribution, we vary the disc inclination, position angle, and temperature power law to fit the data. 

First, we make some considerations to better understand the dependence of the minima on our model parameters in the context of an axisymmetric disc model as outlined above. For a face-on disc model of HD~100546, the only difference between the location of the minima is due to the differences in length between individual baselines. In our case the difference is relatively small, about 1~m, and it introduces a minor effect in shifting the minimum of the 15~m baseline with respect to the 16~m baselines by less than 0.8~$\mu$m towards shorter wavelengths. With increasing disc inclination, the minima of the baselines oriented closest to the minor axis move rapidly to the shorter baselines, and those near the major axis lag behind, as the difference induced by inclination along directions near the disc's major axis is minimal. Therefore, from the location of the minima of baselines near the minor axis of the disc, and from the comparison to baselines  oriented further from the minor axis, we can constrain the inclination and position angle of our science target. In our set of AT measurements, we have three short baselines almost evenly spaced with about  30$\degr$ from one another, as seen in  Fig.~\ref{uvplot}. The strength of correlated fluxes away from the minima (e.g., near 13~$\mu$m) depend on the brightness of the dominant sharp feature, in our case the disc wall and, to a lesser degree, the radial decrease in disc brightness.

\begin{figure}
   \centering
   \includegraphics[width=8cm]{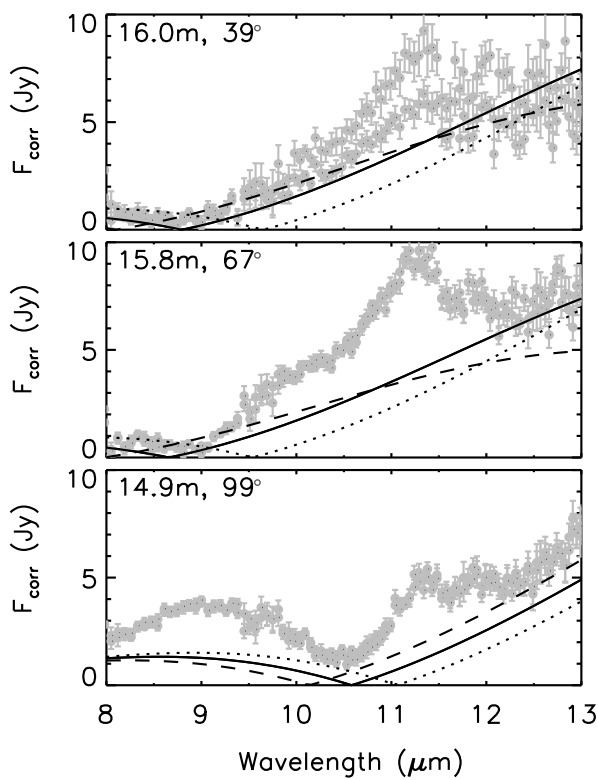}
    \caption{Same data as in Fig.~\ref{corrflux}, but plotted in grey for both calibrations. Full lines show the model prediction for the correlated fluxes on these baselines (see Sect.~\ref{outerdisc}). The dotted and the dashed lines present the model result for inclination angles 45$\degr$ and 61$\degr$ respectively, compared to our model with 53$\degr$ shown with full lines.}
              \label{corrflux2}
    \end{figure}

The fact that the two baselines shown in the upper and middle panels of Fig.~\ref{corrflux} have almost identical correlated fluxes strongly suggests of their similar orientation with respect to the disc minor axis. Any baseline length effects can be excluded due to their essentially same length (16.0 and 15.8~m). Setting the minor axis between these two baselines, at $PA=$145$\degr$ we optimise the fit to the inclination and obtain an inclination of $i=$53$\degr$. Setting a conservative 0.5~$\mu$m tolerance for the location of the minima, our derived values are $PA=$145$\degr \pm$5$\degr$ (measured EoN) and $i=$53$\degr \pm$8$\degr$ (0$\degr$ corresponds to face-on). These values are consistent with those previously derived based on mid-infrared observations \citep[150$\pm$10$\degr$ and 45$\pm$15$\degr$][]{liu2003}, and they compare well to the disc geometry on a larger scale \citep{ardila2007}. \citet{tatulli2011} derive a lower inclination, 33$\pm$11$\degr$, for the rim of the inner disc, which indicates that there may be a slight misalignment with the wall of the outer disc, probed by our observations. However, this may be difficult to corroborate in case of a significant asymmetry (see Sect.~\ref{asym}).
Figure \ref{corrflux} shows how sensitive our data are to the position angle, and Fig.~\ref{corrflux2} shows the sensitivity to the inclination.
 
Finally, with the minima and relative 8~$\mu$m and 13~$\mu$m fluxes at levels consistent with our interferometric data, we require the temperature in the outer disc to match the high 13~$\mu$m fluxes and low 8~$\mu$m fluxes of the short baselines. The fit is obtained for $T_{in,OD}=$230~K and $q_{OD}=$-0.5, as shown in Fig.~\ref{corrflux}. The combined effect of the fit parameters is a bright peak of emission at 11$\pm$1~AU. Degeneracies, in particular between $R_{in,OD}$ and $T_{in,OD}$ are present, but the derived location of the peak brightness is a robust result. 
With this disc model we also roughly match the total fluxes at 13~$\mu$m of $\approx$50~Jy, as seen in Fig.~\ref{flux}. Because of the silicate features and PAH bands arising from the disc on larger scales, the total flux exceeds our model prediction. We explore a range of temperatures and radial distributions of disc brightness, and find that there is a large apparent inconsistency at 8~$\mu$m between the correlated flux levels observed interferometrically and the total flux levels at short wavelengths. A possible scenario is that the excess emission is due to a long wavelength wing of PAH bands centred at 7.7-7.9~$\mu$m, which, as we find in Sect.~\ref{totalflux}, come from a region larger than the UT slit size of 0$\farcs$52. 

\subsection{Asymmetry}\label{asym}

The point symmetry of the brightness of our disc can be tested by the chromatic phase measurements. In case of deviations from point symmetry, the phases exhibit a deviation from zero, since contributions from opposite sides of the star fail to cancel one another completely. Axially symmetric disc models, such as ours, necessarily result in zero phases. Figure~\ref{phase} shows the calibrated chromatic phases of our UT baseline measurements, including the 34~m baseline from 2004 June and the 74~m baseline from 2003 \citep{leinert2004}. Deviations as large as 10-60$\degr$ can be seen on all UT baselines. The precision of these measurements was estimated by cross-calibrating pairs of calibrators and comparing the phase difference for each pair. In this way, we find that the chromatic phases are correct to better than 3.2$\degr$ for the 41~m baselines and 3.8$\degr$ for the 34~m baseline. The bottom panel of Fig.~\ref{phase} shows the difference in the measured phases for two calibrators, HD~102461 and HD~98292, observed at different airmasses. Both calibrators are essentially point sources, and their intrinsic phases correspond to zero (point symmetry). The atmospheric conditions along the line of sight to these calibrators are thus the main contributors to the noise in the phase measurements, and the difference between the two shows these noise levels, varying within $\pm$3$\degr$, as illustrated by the grey area. 

Therefore our data presents clear evidence of a lack of point symmetry in the mid-infrared brightness distribution. Asymmetry persists in our UT data from 2004 June, 2004 December and 2005 December (Fig.~\ref{phase}), as well as in data from even longer, 70~m baseline MIDI observations of \citet{leinert2004} (Fig.~\ref{phase}).

One AT baseline also exhibits clear signatures of asymmetry in the measured phases -- this is the 2006 February 16 dataset (14.9~m, 99$\degr$ baseline), as seen in Fig.~\ref{phase}. Deviations as high as 40$\degr$ are seen in a pronounced, seemingly repetitive pattern along the wavelength. The remaining two AT datasets have very noisy phase and flux measurements in the 8-10.5~$\mu$m range due to low correlated fluxes. For this reason we cannot assess their phases reliably. 

As for UT data, we examined the measured phases of the point source calibrators observed with ATs and find that the atmospheric effects could not amount to more than 5$\%$ phase deviations (Fig.~\ref{phase}). This clearly shows that asymmetries are present on the spatial scales probed by our short AT baselines. Because our AT correlated fluxes are dominated by the bright wall of the disc at 11~AU, we conclude that the asymmetry of the disc wall is the origin of deviations seen in the corresponding phase measurements in the 2006 February 16 AT baseline. The same asymmetry may have influenced the phases seen on the UT baselines. \citet{grady2005a} present the long-slit spectroscopy of HD~100546 in the far ultraviolet suggestive of asymmetry in the brightness profile along the major axis. Thus our data adds to a growing amount of indirect evidence that HD~100546 may host planets or a more massive perturber inside its gap.

\section{Summary}\label{summary}

In this work we have revisited the bright disc around the intermediate-mass star HD~100546. Although plenty of observational data are available for this source, the physical conditions of dust in the areas adjacent to the gap in the planet-forming regions of this disc have been poorly constrained until now. To address this, we took advantage of the VLTI and its ability to observe discs around young stars over a range of spatial scales ranging from the Earth's orbit to the orbit of Uranus. Especially in discs like HD~100546, interferometric observations are the only route to probe the spatial structures near the gap, without being blinded by the star or overwhelmed by disc emission on large scales. The mid-infrared wavelength range covered by the VLTI instrument MIDI is multiply advantageous: 1) The emission is dominated by the disc while stellar contribution is negligible, and 2) The thermal emission of the disc regions of interest is the strongest in mid-infrared compared to longer wavelengths, owing to their temperature range from a few hundred to several hundred Kelvin.

We used 41~m baselines between pairs of UTs (8.2~m) to analyse the inner disc emission. With a suite of shorter, 15~m baselines between pairs of ATs (1.8~m), we probed the transition region, i.e., `disc wall', between the gap and the outer disc. To derive most robust constraints on the radial mid-infrared brightness profile, we assumed black-body emission from featureless dust, and used a very basic model with a gap and radial temperature distributions. We used the observationally constrained parameters from the literature, such as the location of the inner rim (VLTI/AMBER) and approximate location of the wall of the outer disc (VLT/CRIRES, VLT/NACO), and we used the physically consistent disc temperatures from sophisticated radiative transfer modelling of the spectral energy distribution of HD~100546, as guidance in our choice of temperature. 

Our main conclusions can be summarised as follows:
\begin{itemize}
\item{\textit{The inner disc extends no farther than 0.7~AU from the star.} Contrary to prior SED modelling done assuming an inner disc with a 4~AU radius, our uniquely sensitive MIDI data have for the first time provided stringent upper limits to the radial distribution of the dust in the inner disc. We derive a firm upper limit of 0.7~AU. In \cite{mulders2013} we decrease this limit further to 0.3~AU by employing a detailed physical model.}
\item{\textit{The gap is about 10~AU wide and free of detectable mid-infrared emission.} Much larger dust, rocks, and planetesimals are not efficient N-band emitters, and our data do not exclude their presence inside this gap.}
\item{\textit{Our modelling places the radial peak of N-band emission from the outer disc at 11$\pm$1~AU.} This estimate is relatively close to the pronounced peak in ro-vibrational CO line emission observed by \citet{vanderplas2009} and the bright ring of polarised light shown in \citet{quanz2011}. }
\item{\textit{The 7.7~$\mu$m and 8.6~$\mu$m PAH features come from the outer disc.} Absent from our total flux measurements with UTs and ATs, these features arise from an angle that is larger than the slit of our telescopes. On spatial scales, this corresponds to about 100~AU from the star.}
\item{\textit{The mid-infrared brightness distribution of HD~100546 is asymmetrical.} Clear and strong deviations from zero in the chromatic phases in all our UT measurements, including data from \citet{leinert2004}, and also an AT measurement are direct evidence of asymmetry. We exclude the possibility of any atmospheric or instrumental effects causing such phase deviations. Asymmetry persisted in MIDI data collected over the 1.5~year period.}
\end{itemize}

Our results are a useful zeroeth-order model for any sophisticated radiative transfer modelling that wishes to simultaneously study the detailed disc physics and ensure consistency with the interferometric data. This was done in our parallel paper \citet{mulders2013}, allowing constraints on the possible companion from the detailed shape of the disc wall.

The data we present answers several questions, but also opens an avenue toward further investigation. Is the asymmetric spatial distribution of dust in the inner disc long-lived? Follow-up observations on similar baselines with MIDI can answer this. In the case of a long-lived resonant asymmetric structure, how exactly is the dust distributed azimuthally? One of the key standing questions is: Does the inner disc possess sufficient amounts of gas to be viscous and in hydrostatic equilibrium, or has it already evolved to a debris-disc stage in which the dust is large and grinds collisionally to produce the small dust that we observe? If so, HD~100546 would be the first known star to be surrounded by both debris disc in the inner regions and an optically thick disc in the outer regions. The asymmetry seen in the mid-infrared emission of the disc wall at 11~AU is suggestive of dynamical perturbation by one or more bodies, so it warrants further investigations, in particular high-resolution imaging. Future VLTI instrument MATISSE and VLT instrument SPHERE will be able to directly image the planet-forming regions of HD~100546, while the Atacama Millimeter/Submillimeter Array is able to investigate the spatial distribution of large dust grains and molecular gas in the inner tens of AU of this disc. It will be interesting to look for clues to whether the asymmetries we see on small scales and multiple spirals imaged at 100~AU have a common origin.

\begin{acknowledgements}
The research of O. P. leading to these results has received funding from the European Community's Seventh Framework Programme (/FP7/2007-2013/) under grant agreement No 229517. A significant part of the project was completed during the ESO Fellowship of O. P., co-funded by Marie-Curie Actions. O. P. wishes to thank the Max Planck Institute for Astronomy in Heidelberg for the research grant awarded in 2011, which was used for this project. G.D.M. acknowledges support from the Leids Kerkhoven-Bosscha Fonds. The authors thank A. Verhoeff, B. Acke and C. Leinert for sharing their previous data, L.B.F.M Waters and C. A. Hummel for advice and discussions.
\end{acknowledgements}

\bibliography{refs} 

\begin{thebibliography}{50}
\expandafter\ifx\csname natexlab\endcsname\relax\def\natexlab#1{#1}\fi

\bibitem[{{Acke} \& {van den Ancker}(2004)}]{acke2004}
{Acke}, B. \& {van den Ancker}, M.~E. 2004, \aap, 426, 151

\bibitem[{{Acke} \& {van den Ancker}(2006)}]{acke2006}
{Acke}, B. \& {van den Ancker}, M.~E. 2006, \aap, 449, 267

\bibitem[{{Andrews} {et~al.}(2011){Andrews}, {Wilner}, {Espaillat}, {Hughes},
  {Dullemond}, {McClure}, {Qi}, \& {Brown}}]{andrews2011}
{Andrews}, S.~M., {Wilner}, D.~J., {Espaillat}, C., {et~al.} 2011, \apj, 732,
  42

\bibitem[{{Ardila} {et~al.}(2007){Ardila}, {Golimowski}, {Krist}, {Clampin},
  {Ford}, \& {Illingworth}}]{ardila2007}
{Ardila}, D.~R., {Golimowski}, D.~A., {Krist}, J.~E., {et~al.} 2007, \apj, 665,
  512

\bibitem[{{Augereau} {et~al.}(2001){Augereau}, {Lagrange}, {Mouillet}, \&
  {M{\'e}nard}}]{augereau2001}
{Augereau}, J.~C., {Lagrange}, A.~M., {Mouillet}, D., \& {M{\'e}nard}, F. 2001,
  \aap, 365, 78

\bibitem[{{Benisty} {et~al.}(2010){Benisty}, {Tatulli}, {M{\'e}nard}, \&
  {Swain}}]{benisty2010}
{Benisty}, M., {Tatulli}, E., {M{\'e}nard}, F., \& {Swain}, M.~R. 2010, \aap,
  511, A75

\bibitem[{{Boccaletti} {et~al.}(2013){Boccaletti}, {Pantin}, {Lagrange},
  {Augereau}, {Meheut}, \& {Quanz}}]{boccaletti2013}
{Boccaletti}, A., {Pantin}, E., {Lagrange}, A.-M., {et~al.} 2013, ArXiv
  e-prints

\bibitem[{{Bouwman} {et~al.}(2003){Bouwman}, {de Koter}, {Dominik}, \&
  {Waters}}]{bouwman2003}
{Bouwman}, J., {de Koter}, A., {Dominik}, C., \& {Waters}, L.~B.~F.~M. 2003,
  \aap, 401, 577

\bibitem[{{Brown} {et~al.}(2007){Brown}, {Blake}, {Dullemond}, {Mer{\'{\i}}n},
  {Augereau}, {Boogert}, {Evans}, {Geers}, {Lahuis}, {Kessler-Silacci},
  {Pontoppidan}, \& {van Dishoeck}}]{brown2007}
{Brown}, J.~M., {Blake}, G.~A., {Dullemond}, C.~P., {et~al.} 2007, \apjl, 664,
  L107

\bibitem[{{Brown} {et~al.}(2009){Brown}, {Blake}, {Qi}, {Dullemond}, {Wilner},
  \& {Williams}}]{brown2009}
{Brown}, J.~M., {Blake}, G.~A., {Qi}, C., {et~al.} 2009, \apj, 704, 496

\bibitem[{{Calvet} {et~al.}(2002){Calvet}, {D'Alessio}, {Hartmann}, {Wilner},
  {Walsh}, \& {Sitko}}]{calvet2002}
{Calvet}, N., {D'Alessio}, P., {Hartmann}, L., {et~al.} 2002, \apj, 568, 1008

\bibitem[{{Cohen} {et~al.}(1999){Cohen}, {Walker}, {Carter}, {Hammersley},
  {Kidger}, \& {Noguchi}}]{cohen1999}
{Cohen}, M., {Walker}, R.~G., {Carter}, B., {et~al.} 1999, \aj, 117, 1864

\bibitem[{{Crida} {et~al.}(2006){Crida}, {Morbidelli}, \& {Masset}}]{crida2006}
{Crida}, A., {Morbidelli}, A., \& {Masset}, F. 2006, \icarus, 181, 587

\bibitem[{{di Folco} {et~al.}(2009){di Folco}, {Dutrey}, {Chesneau}, {Wolf},
  {Schegerer}, {Leinert}, \& {Lopez}}]{difolco2009}
{di Folco}, E., {Dutrey}, A., {Chesneau}, O., {et~al.} 2009, \aap, 500, 1065

\bibitem[{{Espaillat} {et~al.}(2010){Espaillat}, {D'Alessio}, {Hern{\'a}ndez},
  {Nagel}, {Luhman}, {Watson}, {Calvet}, {Muzerolle}, \&
  {McClure}}]{espaillat2010}
{Espaillat}, C., {D'Alessio}, P., {Hern{\'a}ndez}, J., {et~al.} 2010, \apj,
  717, 441

\bibitem[{{Glindemann} {et~al.}(2003){Glindemann}, {Algomedo}, {Amestica},
  {Ballester}, {Bauvir}, {Bugue{\~n}o}, {Correia}, {Delgado}, {Delplancke},
  {Derie}, {Duhoux}, {di Folco}, {Gennai}, {Gilli}, {Giordano}, {Gitton},
  {Guisard}, {Housen}, {Huxley}, {Kervella}, {Kiekebusch}, {Koehler},
  {L{\'e}v{\^e}que}, {Longinotti}, {M{\'e}nardi}, {Morel}, {Paresce}, {Phan
  Duc}, {Richichi}, {Sch{\"o}ller}, {Tarenghi}, {Wallander}, {Wittkowski}, \&
  {Wilhelm}}]{glindemann2003}
{Glindemann}, A., {Algomedo}, J., {Amestica}, R., {et~al.} 2003, in ESA Special
  Publication, Vol. 522, GENIE - DARWIN Workshop - Hunting for Planets

\bibitem[{{Goto} {et~al.}(2012){Goto}, {van der Plas}, {van den Ancker},
  {Dullemond}, {Carmona}, {Henning}, {Meeus}, {Linz}, \& {Stecklum}}]{goto2012}
{Goto}, M., {van der Plas}, G., {van den Ancker}, M., {et~al.} 2012, \aap, 539,
  A81

\bibitem[{{Grady} {et~al.}(2005{\natexlab{a}}){Grady}, {Woodgate}, {Heap},
  {Bowers}, {Nuth}, {Herczeg}, \& {Hill}}]{grady2005a}
{Grady}, C.~A., {Woodgate}, B., {Heap}, S.~R., {et~al.} 2005{\natexlab{a}},
  \apj, 620, 470

\bibitem[{{Grady} {et~al.}(2005{\natexlab{b}}){Grady}, {Woodgate}, {Bowers},
  {Gull}, {Sitko}, {Carpenter}, {Lynch}, {Russell}, {Perry}, {Williger},
  {Roberge}, {Bouret}, \& {Sahu}}]{grady2005}
{Grady}, C.~A., {Woodgate}, B.~E., {Bowers}, C.~W., {et~al.}
  2005{\natexlab{b}}, \apj, 630, 958

\bibitem[{{Guimar{\~a}es} {et~al.}(2006){Guimar{\~a}es}, {Alencar}, {Corradi},
  \& {Vieira}}]{guimaraes2006}
{Guimar{\~a}es}, M.~M., {Alencar}, S.~H.~P., {Corradi}, W.~J.~B., \& {Vieira},
  S.~L.~A. 2006, \aap, 457, 581

\bibitem[{{Haisch} {et~al.}(2001){Haisch}, {Lada}, \& {Lada}}]{haisch2001}
{Haisch}, Jr., K.~E., {Lada}, E.~A., \& {Lada}, C.~J. 2001, \apjl, 553, L153

\bibitem[{{Hillenbrand}(2008)}]{hillenbrand2008}
{Hillenbrand}, L.~A. 2008, Physica Scripta Volume T, 130, 014024

\bibitem[{{Jaffe}(2004)}]{jaffe2004}
{Jaffe}, W.~J. 2004, in Society of Photo-Optical Instrumentation Engineers
  (SPIE) Conference Series, Vol. 5491, Society of Photo-Optical Instrumentation
  Engineers (SPIE) Conference Series, ed. {W.~A.~Traub}, 715

\bibitem[{{Leinert}(2003)}]{leinert2003b}
{Leinert}, C. 2003, in ESA Special Publication, Vol. 522, GENIE - DARWIN
  Workshop - Hunting for Planets

\bibitem[{{Leinert} {et~al.}(2003){Leinert}, {Graser}, {Przygodda}, {Waters},
  {Perrin}, {Jaffe}, {Lopez}, {Bakker}, {B{\"o}hm}, {Chesneau}, {Cotton},
  {Damstra}, {de Jong}, {Glazenborg-Kluttig}, {Grimm}, {Hanenburg}, {Laun},
  {Lenzen}, {Ligori}, {Mathar}, {Meisner}, {Morel}, {Morr}, {Neumann}, {Pel},
  {Schuller}, {Rohloff}, {Stecklum}, {Storz}, {von der L{\"u}he}, \&
  {Wagner}}]{leinert2003a}
{Leinert}, C., {Graser}, U., {Przygodda}, F., {et~al.} 2003, \apss, 286, 73

\bibitem[{{Leinert} {et~al.}(2004){Leinert}, {van Boekel}, {Waters},
  {Chesneau}, {Malbet}, {K{\"o}hler}, {Jaffe}, {Ratzka}, {Dutrey}, {Preibisch},
  {Graser}, {Bakker}, {Chagnon}, {Cotton}, {Dominik}, {Dullemond},
  {Glazenborg-Kluttig}, {Glindemann}, {Henning}, {Hofmann}, {de Jong},
  {Lenzen}, {Ligori}, {Lopez}, {Meisner}, {Morel}, {Paresce}, {Pel},
  {Percheron}, {Perrin}, {Przygodda}, {Richichi}, {Sch{\"o}ller}, {Schuller},
  {Stecklum}, {van den Ancker}, {von der L{\"u}he}, \& {Weigelt}}]{leinert2004}
{Leinert}, C., {van Boekel}, R., {Waters}, L.~B.~F.~M., {et~al.} 2004, \aap,
  423, 537

\bibitem[{{Liu} {et~al.}(2003){Liu}, {Hinz}, {Meyer}, {Mamajek}, {Hoffmann}, \&
  {Hora}}]{liu2003}
{Liu}, W.~M., {Hinz}, P.~M., {Meyer}, M.~R., {et~al.} 2003, \apjl, 598, L111

\bibitem[{{Mathar}(2007)}]{mathar2007}
{Mathar}, R.~J. 2007, Baltic Astronomy, 16, 287

\bibitem[{{Morel} {et~al.}(2004){Morel}, {Ballester}, {Bauvir}, {Biereichel},
  {Cuby}, {Galliano}, {Haddad}, {Housen}, {Hummel}, {Kaufer}, {Kervella},
  {Percheron}, {Puech}, {Rantakyro}, {Richichi}, {Sabet}, {Schoeller},
  {Spyromilio}, {Vannier}, {Wallander}, {Wittkowski}, {Leinert}, {Graser},
  {Neumann}, {Jaffe}, \& {de Jong}}]{morel2004}
{Morel}, S., {Ballester}, P., {Bauvir}, B., {et~al.} 2004, in Society of
  Photo-Optical Instrumentation Engineers (SPIE) Conference Series, Vol. 5491,
  Society of Photo-Optical Instrumentation Engineers (SPIE) Conference Series,
  ed. {W.~A.~Traub}, 1666

\bibitem[{{Mulders} {et~al.}(2013){Mulders}, {Paardekooper}, {Pani{\'c}},
  {Dominik}, {van Boekel}, \& {Ratzka}}]{mulders2013}
{Mulders}, G.~D., {Paardekooper}, S.-J., {Pani{\'c}}, O., {et~al.} 2013, \aap,
  557, A68

\bibitem[{{Mulders} {et~al.}(2011){Mulders}, {Waters}, {Dominik}, {Sturm},
  {Bouwman}, {Min}, {Verhoeff}, {Acke}, {Augereau}, {Evans}, {Henning},
  {Meeus}, \& {Olofsson}}]{mulders2011}
{Mulders}, G.~D., {Waters}, L.~B.~F.~M., {Dominik}, C., {et~al.} 2011, \aap,
  531, A93

\bibitem[{{Paardekooper} \& {Mellema}(2004)}]{paardekooper2004}
{Paardekooper}, S.-J. \& {Mellema}, G. 2004, \aap, 425, L9

\bibitem[{{Pani{\'c}} {et~al.}(2010){Pani{\'c}}, {van Dishoeck}, {Hogerheijde},
  {Belloche}, {G{\"u}sten}, {Boland}, \& {Baryshev}}]{panic2010}
{Pani{\'c}}, O., {van Dishoeck}, E.~F., {Hogerheijde}, M.~R., {et~al.} 2010,
  \aap, 519, A110

\bibitem[{{Pantin} {et~al.}(2000){Pantin}, {Waelkens}, \&
  {Lagage}}]{pantin2000}
{Pantin}, E., {Waelkens}, C., \& {Lagage}, P.~O. 2000, \aap, 361, L9

\bibitem[{{Pi{\'e}tu} {et~al.}(2006){Pi{\'e}tu}, {Dutrey}, {Guilloteau},
  {Chapillon}, \& {Pety}}]{pietu2006}
{Pi{\'e}tu}, V., {Dutrey}, A., {Guilloteau}, S., {Chapillon}, E., \& {Pety}, J.
  2006, \aap, 460, L43

\bibitem[{{Quanz} {et~al.}(2013){Quanz}, {Amara}, {Meyer}, {Kenworthy},
  {Kasper}, \& {Girard}}]{quanz2013}
{Quanz}, S.~P., {Amara}, A., {Meyer}, M.~R., {et~al.} 2013, \apjl, 766, L1

\bibitem[{{Quanz} {et~al.}(2011){Quanz}, {Schmid}, {Geissler}, {Meyer},
  {Henning}, {Brandner}, \& {Wolf}}]{quanz2011}
{Quanz}, S.~P., {Schmid}, H.~M., {Geissler}, K., {et~al.} 2011, \apj, 738, 23

\bibitem[{{Quillen}(2006)}]{quillen2006}
{Quillen}, A.~C. 2006, \apj, 640, 1078

\bibitem[{{Ratzka} \& {Leinert}(2005)}]{ratzka2005}
{Ratzka}, T. \& {Leinert}, C. 2005, Astronomische Nachrichten, 326, 570

\bibitem[{{Ratzka} {et~al.}(2007){Ratzka}, {Leinert}, {Henning}, {Bouwman},
  {Dullemond}, \& {Jaffe}}]{ratzka2007}
{Ratzka}, T., {Leinert}, C., {Henning}, T., {et~al.} 2007, \aap, 471, 173

\bibitem[{{Ratzka} {et~al.}(2009){Ratzka}, {Leinert}, {van Boekel}, \&
  {Schegerer}}]{ratzka2009}
{Ratzka}, T., {Leinert}, C., {van Boekel}, R., \& {Schegerer}, A.~A. 2009, in
  Science with the VLT in the ELT Era, ed. {A.~Moorwood}, 101

\bibitem[{{Tatulli} {et~al.}(2011){Tatulli}, {Benisty}, {M{\'e}nard},
  {Varni{\`e}re}, {Martin-Za{\"i}di}, {Thi}, {Pinte}, {Massi}, {Weigelt},
  {Hofmann}, \& {Petrov}}]{tatulli2011}
{Tatulli}, E., {Benisty}, M., {M{\'e}nard}, F., {et~al.} 2011, \aap, 531, A1

\bibitem[{{Thalmann} {et~al.}(2010){Thalmann}, {Grady}, {Goto}, {Wisniewski},
  {Janson}, {Henning}, {Fukagawa}, {Honda}, {Mulders}, {Min},
  {Moro-Mart{\'{\i}}n}, {McElwain}, {Hodapp}, {Carson}, {Abe}, {Brandner},
  {Egner}, {Feldt}, {Fukue}, {Golota}, {Guyon}, {Hashimoto}, {Hayano},
  {Hayashi}, {Hayashi}, {Ishii}, {Kandori}, {Knapp}, {Kudo}, {Kusakabe},
  {Kuzuhara}, {Matsuo}, {Miyama}, {Morino}, {Nishimura}, {Pyo}, {Serabyn},
  {Shibai}, {Suto}, {Suzuki}, {Takami}, {Takato}, {Terada}, {Tomono}, {Turner},
  {Watanabe}, {Yamada}, {Takami}, {Usuda}, \& {Tamura}}]{thalmann2010}
{Thalmann}, C., {Grady}, C.~A., {Goto}, M., {et~al.} 2010, \apjl, 718, L87

\bibitem[{{van Boekel}(2004)}]{vanboekelPhD}
{van Boekel}, R. 2004, PhD thesis, FNWI: Sterrenkundig Instituut Anton
  Pannekoek, Postbus 19268, 1000 GG Amsterdam, The Netherlands

\bibitem[{{van Boekel} {et~al.}(2004){van Boekel}, {Waters}, {Dominik},
  {Dullemond}, {Tielens}, \& {de Koter}}]{vanboekel2004}
{van Boekel}, R., {Waters}, L.~B.~F.~M., {Dominik}, C., {et~al.} 2004, \aap,
  418, 177

\bibitem[{{van den Ancker} {et~al.}(1997){van den Ancker}, {The}, {Tjin A
  Djie}, {Catala}, {de Winter}, {Blondel}, \& {Waters}}]{vandenancker1997}
{van den Ancker}, M.~E., {The}, P.~S., {Tjin A Djie}, H.~R.~E., {et~al.} 1997,
  \aap, 324, L33

\bibitem[{{van der Plas} {et~al.}(2009){van der Plas}, {van den Ancker},
  {Acke}, {Carmona}, {Dominik}, {Fedele}, \& {Waters}}]{vanderplas2009}
{van der Plas}, G., {van den Ancker}, M.~E., {Acke}, B., {et~al.} 2009, \aap,
  500, 1137

\bibitem[{{van Leeuwen}(2007)}]{vanleeuwen2007}
{van Leeuwen}, F., ed. 2007, Astrophysics and Space Science Library, Vol. 350,
  {Hipparcos, the New Reduction of the Raw Data}

\bibitem[{{Verhoeff}(2009)}]{verhoeffPhD}
{Verhoeff}, A. 2009, PhD thesis, Sterrenkundig Instituuut ''Anton Pannekoek'',
  University of Amsterdam

\bibitem[{{Verhoelst}(2005)}]{verhoelstPhD}
{Verhoelst}, T. 2005, PhD thesis, Institute of Astronomy, K.U.Leuven, Belgium

\end{thebibliography}
\end{document}